\documentclass[cits]{PoS}

\title{Simulations of the origin and fate of the Galactic Center cloud G2}

\ShortTitle{Simulations of the origin and fate of the Galactic Center cloud G2}

\author{\speaker{Marc Schartmann}$^{a,b}$, Andreas Burkert$^{a,b}$\thanks{Max Planck Fellow.}, Christian Alig$^{a,b}$,
        Stefan Gillessen$^{b}$, Reinhard Genzel$^{b}$, Frank Eisenhauer$^{b}$, Tobias Fritz$^{b}$, Alessandro Ballone$^{a,b}$\\
       \llap{$^a$}Universit\"atssternwarte M\"unchen, Scheinerstr. 1, 81679 M\"unchen\\
       \llap{$^b$}Max-Planck-Institut f\"ur extraterrestrische Physik, Postfach 1312, Giessenbachstr., 85741 Garching\\
       E-mail: \email{ schartmann@mpe.mpg.de}}

\abstract{We investigate the origin and fate of the recently discovered gas cloud G2 close to the Galactic Center.
Our hydrodynamical simulations focussing on the dynamical evolution of the cloud in combination with currently available 
observations favour two scenarios: a {\it Compact Cloud} which started around the year 1995 and an extended {\it Spherical Shell} of gas, 
with an apocenter distance within the disk(s) of young stars. The former is able to explain the detected signal of G2 in the 
position-velocity-diagram of the year 2008.5 and 2011.5 data. The latter can account for both, G2's signal as well as the 
fainter extended tail-like structure G2t seen at larger distances to the black hole and smaller velocities.
From these first idealised simulations we expect a rise of the current 
activity of Sgr A* shortly after the closest approach and a constant feeding through a nozzle-like structure over a long period.  
The near future evolution of the cloud will be a sensitive probe of the conditions of the gas distribution in the milli-parsec 
environment of the massive black hole in the Galactic Center and will also give us invaluable information of the feeding of black 
holes and the activation of the central source.}

\FullConference{Nuclei of Seyfert galaxies and QSOs - Central engine \& conditions of star formation\\
		 November 6-8, 2012\\
		 Max-Planck-Insitut f\"ur Radioastronomie (MPIfR), Bonn, Germany}

\begin{document}

\section{Introduction}

In the last years, the Galactic Center region has become an even more exciting laboratory for the study of physical processes 
under extreme conditions: A fast-moving gas cloud was discovered in the radial regime of the S-stars \cite{Gillessen_12}. 
Including data from 2002 to 2012, a very accurate orbit was determined \cite{Gillessen_13}.    
The cloud follows a highly eccentric trajectory (e=0.97), bringing it to a distance from the supermassive black hole (SMBH) of only
2200 Schwarzschild radii, which will be reached in late 2013. 
The observed luminosity together with case B recombination theory allows an estimate of the gas mass of the cloud, which results in 
$1.7 \cdot 10^{28}$g or roughly 3 earth masses.

\section{Two basic hydrodynamical models}

The idea of these first hydrodynamical simulations was to set up two basic models of diffuse clouds. 
The so-called {\it Compact Cloud} was supposed to be able 
to describe the observations of the actual G2 
component and to follow the observed orbit for this component. It is a cloud, which started its journey in 1995 on 
the orbit initially in pressure equilibrium. This seems to be a reasonable assumption as we analytically estimated that overpressured clouds will 
quickly adjust to pressure equilibrium, whereas highly underpressured clouds would probably quickly collapse and fragment into pieces \cite{Burkert_12}. 
Hence the assumption of pressure equilibrium at least in the region around the apo-centre seems to be reasonable. 
The so-called {\it Spherical Shell} was supposed to describe both components, G2 (head) and G2t (tail). Hence it must be a larger cloud complex and G2 can 
then be interpreted as the front part of this cloud. It can have a higher mass and the centre of mass of the cloud can be on a 
slightly different orbit.

\section{Setup of the hydrodynamical models}

The most important ingredient for the modelling - since we are currently witnessing the tidal disruption - is 
the gravitation of the SMBH. Its mass is known to a high accuracy 
and was determined to be $4.3\cdot 10^6\,\mathrm{M}_\odot$ from fits to S-star orbits \cite{Gillessen_09} and 
is modelled as a Newtonian point potential. The cloud is embedded into a hot and dense 
magnetized accretion flow.
In our simulations, we describe it by an ADAF ({\it Advection Dominated Accretion Flow}) model of \cite{Yuan_03}. 
With these ingredients, we solve the Euler equations with the help of the {\sc PLUTO} \cite{Mignone_07} code in a 2D 
Cartesian coordinate system, using an isothermal equation of state for the cloud, as the thermodynamics 
of the cloud is supposed to be given by the ionizing radiation of the young stars surrounding Sgr A*. 
For details on the numerical treatment, we refer to \cite{Schartmann_12}.

\section{Evolution of the density distribution}

\begin{figure} 
\includegraphics[width=.95\textwidth]{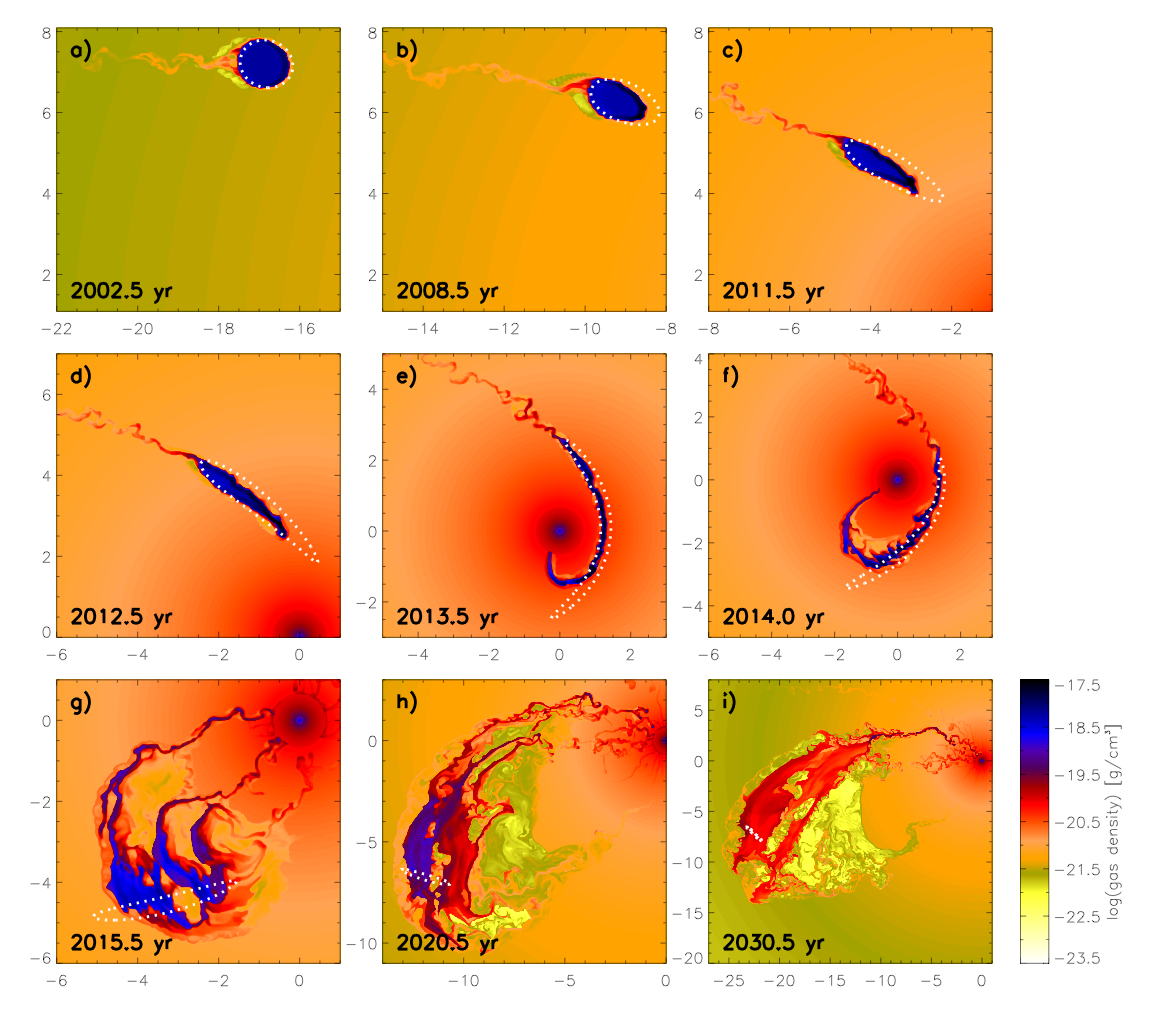} 
\caption{Density evolution of the {\it Compact Cloud} scenario. The dotted lines denote the limiting contours of a test particle simulation. 
         Labels are given in milli-parsec.} 
\label{fig1} 
\end{figure} 

Fig.~\ref{fig1} shows the time evolution of our {\it Compact Cloud} standard model. The white dotted line depicts the evolution of a test particle simulation, 
which indicates that the early evolution (a-d) is dominated by tidal interaction with the black hole and the cloud develops an elongated structure. Additionally 
the effect of the ram pressure of the ambient medium can be clearly seen, compressing the front part of the cloud, which is falling behind the ballistic 
orbits. Furthermore, it leads to a hydrodynamically stripped tail, which is dispersed by Kelvin-Helmholtz-instabilities, 
but with very low mass-loss rates. Close to peri-centre passage (panel e), the cloud has transformed into a long and thin filament, showing a meandering 
shape. The latter is caused by the beginning growth of the Kelvin-Helmholtz instability at the edges of the filamentary cloud. 
Due to the increasing density of the ambient medium towards the SMBH, the leading part of the cloud is able to exchange angular momentum with the ambient medium, 
which leads to accretion towards the centre. Being a boundary effect, this happens in a nozzle-like feature. 
However, the ram pressure interaction is not enough to capture the whole cloud on short timescale, but the dominant fraction 
keeps following the original orbit. Right after the peri-centre passage the cloud contracts again, mainly due to gravitational effects. However, 
the cloud has also increased in size dramatically, which leads to a significant interaction of the upstream part of the cloud with the ambient 
medium and cloud gas is fed towards the centre within the nozzle like feature. 
Concerning the {\it Spherical Shell} scenario, the basic behaviour is quite similar: 
It starts in spherical shape at its apocentre distance and soon develops the typical drop-like structure due to the tidal interaction with the central black 
hole. A tail of hydrodynamically stripped material is formed as well. As the time evolution from its apocentre is much longer, it develops the typical 
disturbances due to Kelvin-Helmholtz-instabilities invoked by the shear flow at the cloud's boundaries. Hence, parts of the cloud reach even closer 
to the centre, where density and ram pressure effects are higher. Therefore, this simulation leads to a rather filamentary disk-like configuration and 
much higher mass transfer rates towards the centre at earlier times (see discussion in Sect.~\ref{sec:conn_obs_em}).

\section{Connection to observations: the PV diagram}
\label{sec:conn_obs_pv}

\begin{figure}[b]
\includegraphics[width=.95\textwidth]{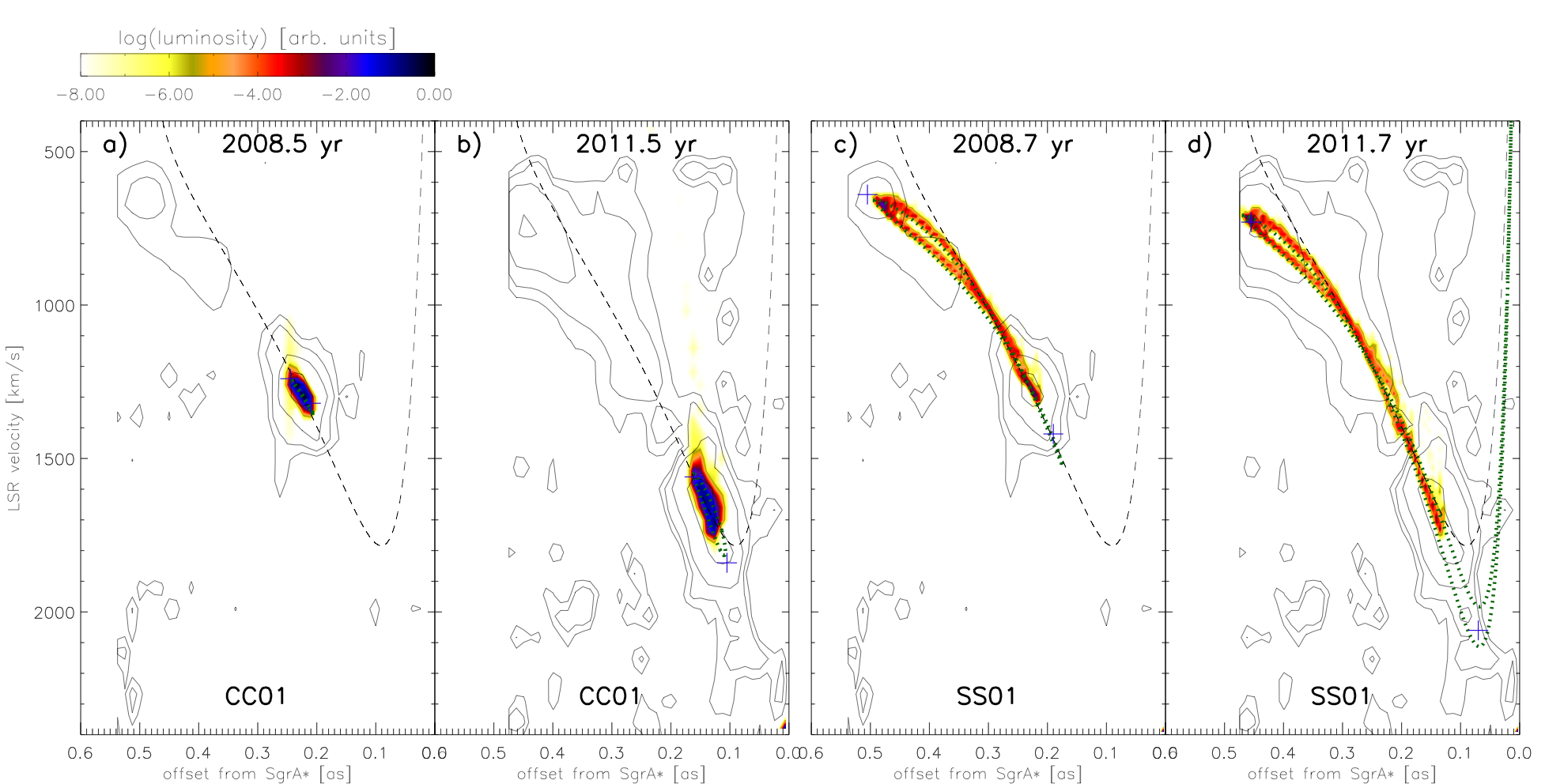} 
\caption{Comparison of the position-velocity-diagrams for the {\it Compact Cloud} model and the {\it Spherical Shell} model with observations (background contours).
The distance to Sgr A* (projected on the sky) is plotted against the line-of-sight velocity. Dashed lines show the evolution of a test particle on the observed orbit
and dotted lines represent test particle simulations.} 
\label{fig2} 
\end{figure} 

In the following, we will try to connect our simulations to observable quantities. This can first of all be done with the position-velocity-diagram (Fig.~\ref{fig2}),
where the line-of-sight velocity is plotted against the distance of the cloud 
to the SMBH as projected on the sky. The observations are given as the background contours and clearly show the two-component structure discussed 
above: G2, the high surface brightness compact component and a conically shaped trailing component, G2t. 
Concentrating on G2, a shearing of the cloud between the two epochs is visible as well in velocity as in position space, 
which means that we are currently witnessing the tidal disruption of the gas cloud. 
The dashed line is the complete evolution of the observed orbit and the various panels show the PV diagrams for the {\it Compact Cloud} model in 2008 and 2011
(panel a and b) and for the {\it Spherical Shell} scenario for the same times (panel c and d).
As expected, the hydro simulations more or less follow the evolution of the collisionless test particles (dotted lines) up to close to the peri centre passage. 
Hydrodynamical effects become most important during and after the peri centre passage, which finally leads to the disruption of the cloud. 
Hence, huge changes are expected within 2013. Our {\it Compact Cloud} model can quite 
accurately describe both of the observations of the G2 component and for the case of the {\it Spherical Shell} also the tail component. 
This is by construction, as we were interested in possible initial conditions of the two diffuse cloud scenarios. 
These models now enable us to make predictions for the near-future evolution, which can be directly compared to upcoming observations.
Remarkably, the two models differ significantly: much higher line-of-sight velocities are reached for the case of the {\it Spherical Shell} model. 
This enables us to rule out several models and adjust parameters in order to get a better idea of the origin of the cloud.

\section{Connection to observations: the EM signal}
\label{sec:conn_obs_em}

\begin{figure}[b]
\includegraphics[width=.6\textwidth]{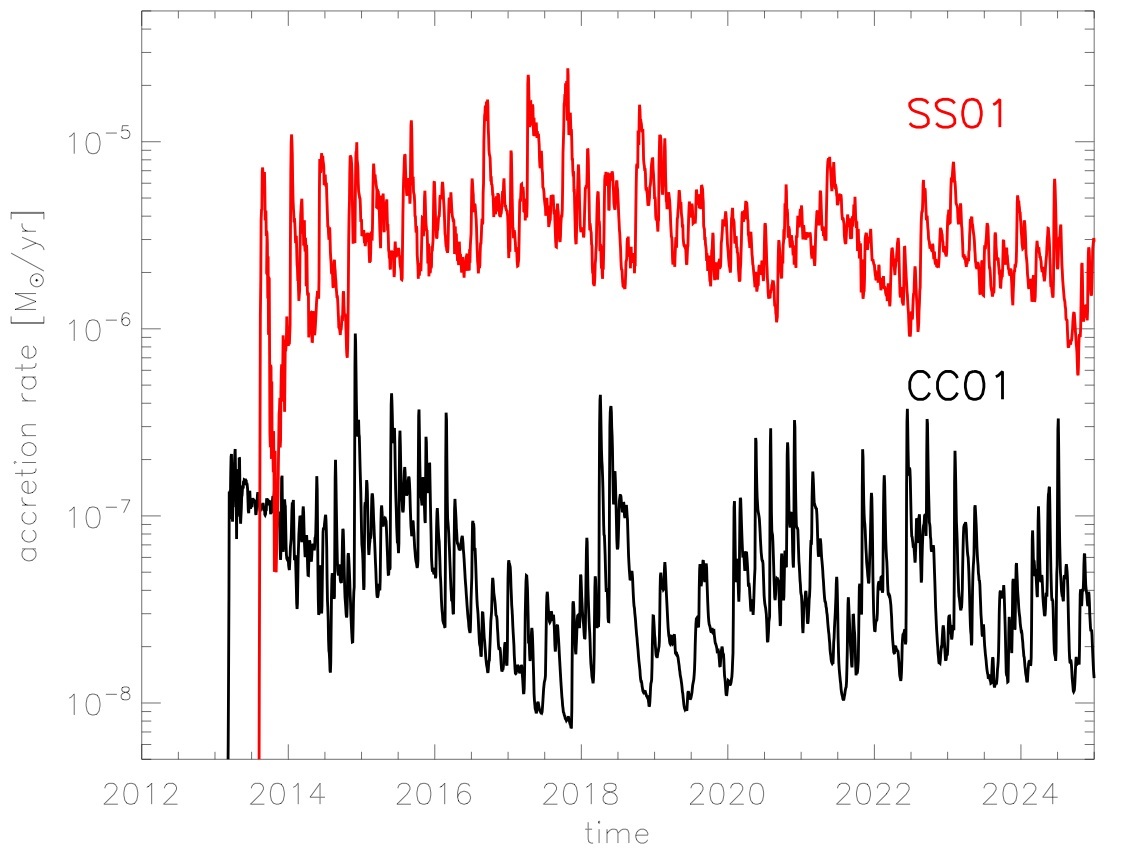} 
\caption{Mass accretion rates through the inner boundary of the computational domain for the {\it Compact Cloud} scenario (black line) and the
         {\it Spherical Shell} scenario (red line).} 
\label{fig3} 
\end{figure} 

Another possibility to connect our simulations to observations is via estimating the change in the electromagnetic signal arising from the 
additional accretion of gas towards the SMBH. This is a non-trivial task and we  
only present a very first approach. Fig.~\ref{fig3} shows the time resolved accretion rate through the inner boundary of our domain ($2\cdot 10^{15}$cm) 
for the two standard models. The accretion rates are quite moderate and hence the timescale to swallow the cloud is quite large. 
It takes roughly until 2060 to swallow 40\% of the mass in the {\it Compact Cloud} scenario. 
Accretion rates are much higher for the {\it Spherical Shell} model, as the initial mass of this model is about a factor of 30 higher 
compared to the {\it Compact Cloud}. But the behaviour otherwise is quite similar, also showing large variations spanning roughly an order of magnitude.  
To translate these time varying accretion rates into observable luminosities in the X-ray and infrared wavebands, we assume that the cold cloud will be 
dispersed and heat up to join the hot accretion flow. We then use ADAF models \cite{Yuan_04}, for which the emitted spectra have been calculated as 
a function of accretion rate and integrate over the respective wavebands. Quite interestingly as well, this does not lead to a significantly boosted 
mean emission for the {\it Compact Cloud} model in comparison to the quiescent value, 
but to boost factors of 20 in the X-rays and 70 in the IR for the {\it Spherical Shell} model. Hence this would 
be another possibility to distinguish between our two favourite models.
For more details on the modelling and simulations we refer to \cite{Burkert_12,Schartmann_12}.

\section{Discussion and conclusions}
Recently a dusty ionized gas cloud was discovered on its way towards the SMBH in our Galactic Centre on a very eccentric orbit. 
In these proceedings, we present idealised hydrodynamical simulations in order to understand where the cloud might have originated and what its  
fate will be in the coming years. This resulted 
in two possible models: a {\it Compact Cloud} starting in the year 1995 and a {\it Spherical Shell}, which started in apocentre in 1927. 
The latter is currently favoured by us due to the following reasons:
(i) It is able to explain both, the G2 (head) as well as the G2t (tail) component. 
(ii) It enables a starting point within the disks of young stars, which are able to provide enough mass, whereas we were unable to 
identify a significant source of mass close to the initial position of the {\it Compact Cloud} model. 
(iii) A {\it Spherical Shell} is a typical
structure of a stellar wind interacting with the ambient medium and additionally the tail is difficult to explain in any other scenario. 
Our simulations show that the evolution of the cloud is a sensitive probe of the environment in the Galactic Centre region. 
Hence the next few years will be very exciting as we might be able to directly probe the immediate vicinity of a SMBH. 
In future, detailed hydrodynamical simulations of physical mechanisms (e.~g.~stellar winds) are needed in order to interpret the observations
and what we have learned so far from our phenomenological approach.

\acknowledgments{This work was supported by the Deutsche Forschungsgemeinschaft
priority program 1573 (``Physics of the Insterstellar Medium''). 
Computer resources for this project have been provided by the 
Leibniz Supercomputing Center under grant: h0075.}

\end{document}